\documentstyle[namedreferences,psfig]{spse2006}

\renewcommand{\thepage}{}   %% no page numbering; we do that

\def\opencite#1{\citeauthor{#1}, \citeyear{#1}} 
\renewcommand{\thepage}{}   %% no page numbering; we do that

\runningtitle{Solar Research Programs at IRSOL}  %% for too long a title
\begin{opening}
  \title{Solar Research Programs at IRSOL, Switzerland}  	    %% will be capitalized
  \author{Renzo Ramelli}
  \author{Michele Bianda\thanks{Also affiliated with ETH Zurich, CH-8092
  Zurich, Switzerland}}
  \institute{Istituto Ricerche Solari Locarno, Via Patocchi~57,
             CH-6005, Locarno, Switzerland}
  \author{Jan O. Stenflo\thanks{also affiliated with the Faculty of Mathematics \& Science, University of Zurich}}
  \institute{Institute of Astronomy, ETH Zentrum, CH-8092 Zurich,
             Switzerland}
  \author{P. Jetzer}
    \institute{Institute for Theoretical Physics,
University of Z\"urich, Winterthurerstrasse 190,
CH-8057 Z\"urich, Switzerland}
\end{opening}

\begin{document}
\begin{abstract}
The Zurich IMaging POLarimeter (ZIMPOL) developed at ETH Zurich and
installed permanently at the Gregory Coud\'e Telescope at
Istituto Ricerche Solari Locarno (IRSOL) allows a
polarimetric precision down to 10$^{-5}$ to be reached. This makes it possible
to perform several accurate spectro-polarimetric
measurements of scattering polarization and to investigate solar
magnetic fields through the signatures of the Hanle and Zeeman effects. 
The research programs are currently being extended to monochromatic imaging of the Stokes vector 
with a recently installed Fabry-Perot rapidly tunable filter
system with a narrow pass band of about 30~m\AA.
The spatial resolution is being improved by the installation of an
Adaptive Optics system.

\end{abstract}

%%%%%%%%%%%%%%%%%%%%%%%%%%%%%%%%%%%%%%%%%%%%%%%%%%%%% INTRODUCTION
\section{Introduction}

The great advances in high precision polarimetry that have been
achieved with the introduction of the Zurich IMaging POLarimeter
(ZIMPOL) a decade ago opened a new window in solar physics.
Polarimetry is in fact a very powerful tool that can be used 
to study solar magnetic fields as well as the physical
processes behind the generation of polarization  
in atomic and molecular spectral lines. 
Magnetic field measurements through Zeeman effect
signatures, which appear in the presence of 
strong and oriented magnetic fields,
have long been performed at many observatories. 
With the high polarimetric precision of ZIMPOL it has become 
possible to extend the magnetic field diagnostics to weak fields and
to fields which are tangled on scales below the spatial resolution, 
which are invisible to the Zeeman effect but get revealed 
by the Hanle effect \cite{hanle} (for details see \opencite{javier06}).

Spectro-polarimetry is currently the main field of research at the Istituto
Ricerche Solari Locarno (IRSOL). Advantage is taken from the circumstance that
a ZIMPOL system is permanently installed at IRSOL. In
addition the Gregory Coud\'e Telescope (GCT) of the observatory is
very well suited for polarimetric measurements, since the amount of
instrumental polarization is low and stays practically constant
during the observing day, since it is a function of declination
only. Therefore it can easily be accounted for.
With a Fabry-Perot filter system and an adaptive optics system
recently installed at IRSOL we plan to start
several new interesting projects.

\section{IRSOL - The Institute}

\begin{figure}[t!]
  \centerline{\psfig{figure=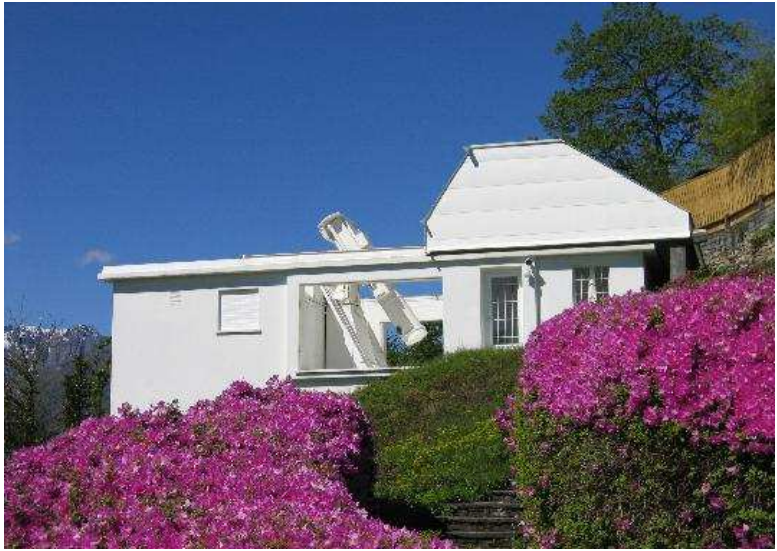,width=0.8\linewidth}}
  \caption[]{\label{osserv}
IRSOL Observatory. In the left part of the building there is the observing room, 
while the spectrograph room is underground. All is painted white with 
titanium dioxide. The institute is located 500 m above sea-level, above 
the city of Locarno, Switzerland.
}
\end{figure}

The observatory at the Istituto Ricerche Solari Locarno (IRSOL)
(Figure \ref{osserv}), located in southern Switzerland, was
constructed in 1960 by the Universit\"ats-Sternwarte G\"ottingen (USG), 
Germany. In 1984, after USG moved its observing activity to the
facilities at Observatorio del Teide on Tenerife, a local foundation
(FIRSOL) acquired the observatory in Locarno. The partially dismantled 
instrumentation was rebuilt and improved, 
in collaboration with USG, University of 
Applied Sciences of Wiesbaden (Germany), and the  
Institute of Astronomy at ETH Zurich.
The scientific collaboration with ETH Zurich
allowed the implementation at IRSOL of an important polarimetry 
observing program, first with a beam exchange polarimeter and then with ZIMPOL.

\section{Instrumentation at IRSOL}

\begin{figure}[htb]
  \centerline{\psfig{figure=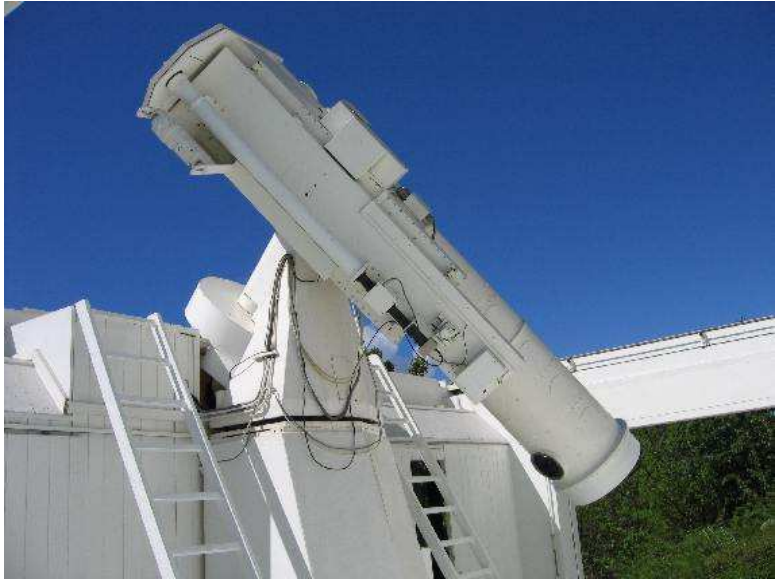,width=0.8\linewidth}}
  \caption[]{\label{telescope}The 45 cm aperture Gregory Coud\'e vacuum telescope.
}
\end{figure}

\begin{figure}[htb]
  \centerline{\psfig{figure=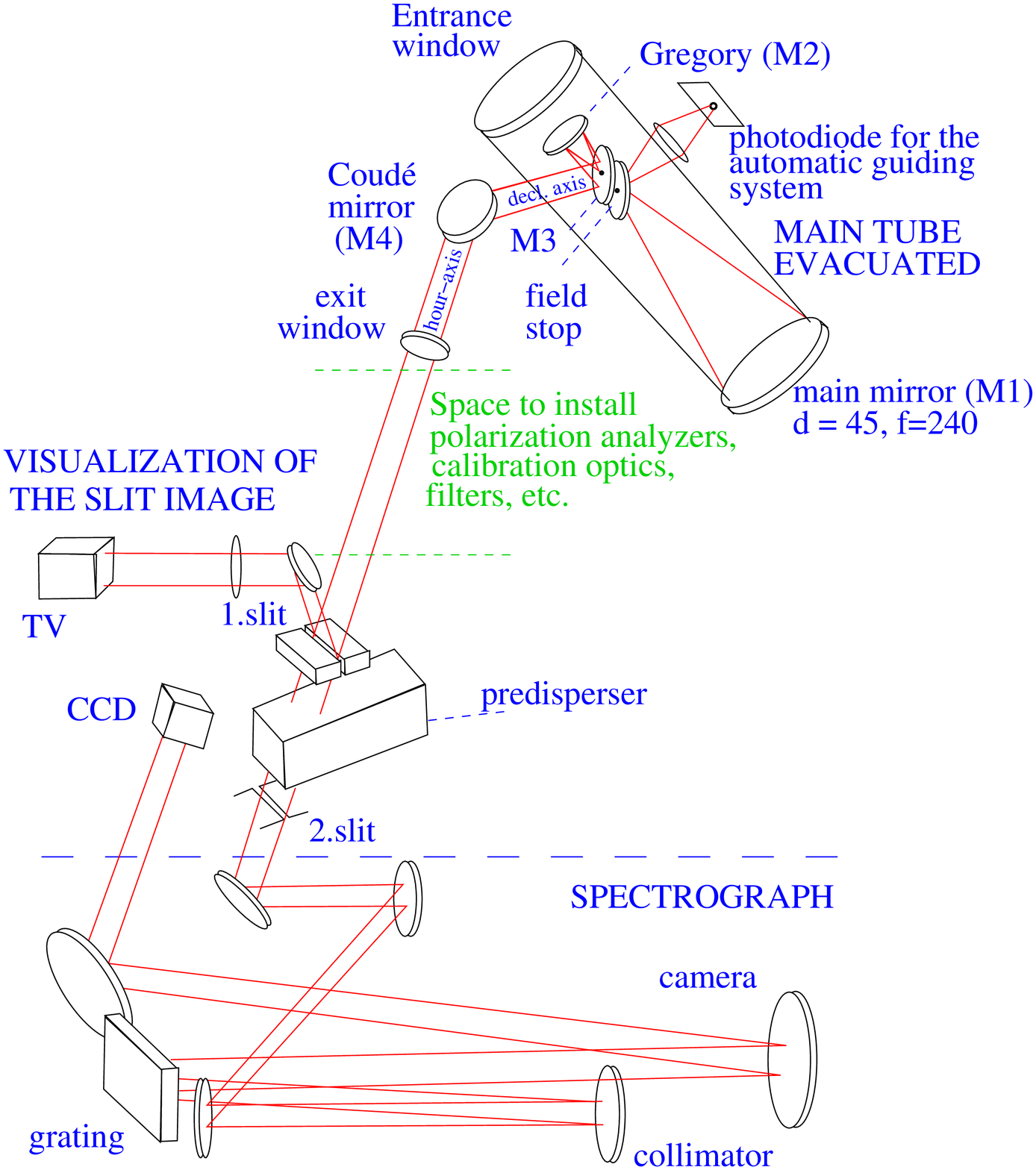,width=0.9\linewidth}}
  \caption[]{\label{tel-scheme}Optical scheme of the Gregory
    Coud\'e Telescope with the spectrograph.
}
\end{figure}

The IRSOL telescope (Figure \ref{telescope} and \ref{tel-scheme}) is a 45~cm aperture 
Gregory Coud\'e  with 24~m effective focal length. 
The field stop at the prime focus reduces the field of view
to a 200 arcsec diameter circular image. The rest of the solar image
is reflected away from the main light beam. 
This reduces heating and scattered light and is of particular advantage when observing low intensity structures 
like sunspots, spicules and prominences. 
The relative orientation of the two folding mirrors M3 and M4 (Coud\'e) changes only with declination and 
is orthogonal at the time of the equinoxes. As a consequence the instrumental 
polarization, originating through oblique reflections, is almost 
constant during the day and virtually vanishes during the equinoxes \cite{sanchez91}. 
A Gregory Coud\'e telescope is thus very well suited for 
polarimetric measurements. 

An automatic guiding system developed by the University of
Applied Sciences Wiesbaden \cite{gerd} is also available. Its
operation is based on the solar image obtained from the light rejected 
by the field stop at the primary focus.

The Czerny-Turner spectrograph with 10 m focal length is based on a 
180 $\times$ 360 mm grating with 300 lines per mm and 63$^{\circ}$ Blaze angle. A 
prism based predisperser allows to select the 
spectral band entering in the 
spectrograph without overlap of the grating orders.

\begin{figure}[htb]
  \centerline{\psfig{figure=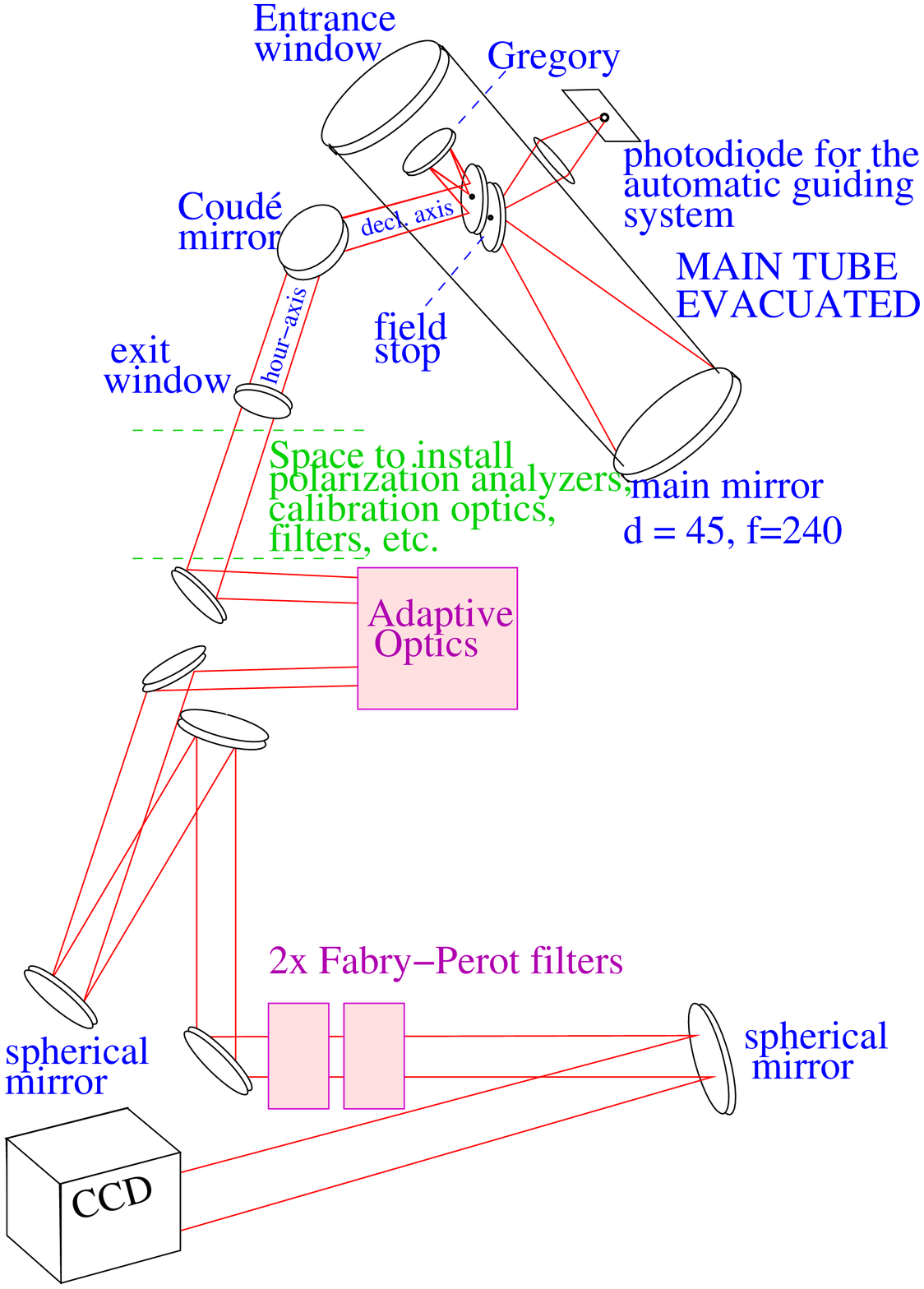,width=0.8\linewidth}}
  \caption[]{\label{fp-scheme}Optical scheme of the Gregory
    Coud\'e Telescope for measurements with the Fabry-Perot filter system.
}
\end{figure}

Monochromatic imaging observations of the solar surface can be performed using 
the recently installed Fabry-Perot filter system in collimated configuration 
\cite{alex} (Figure \ref{fp-scheme}). The system uses two 
temperature controlled lithium niobate etalons with an aperture 
of 70~mm. The transmitted wavelength can be selected by electrically tuning 
the refractive index of the etalon medium, by varying the 
temperature, or by tilting the etalon. The bandwidth is about 30~m\AA. 

\begin{figure}[htb]
  \centerline{\psfig{figure=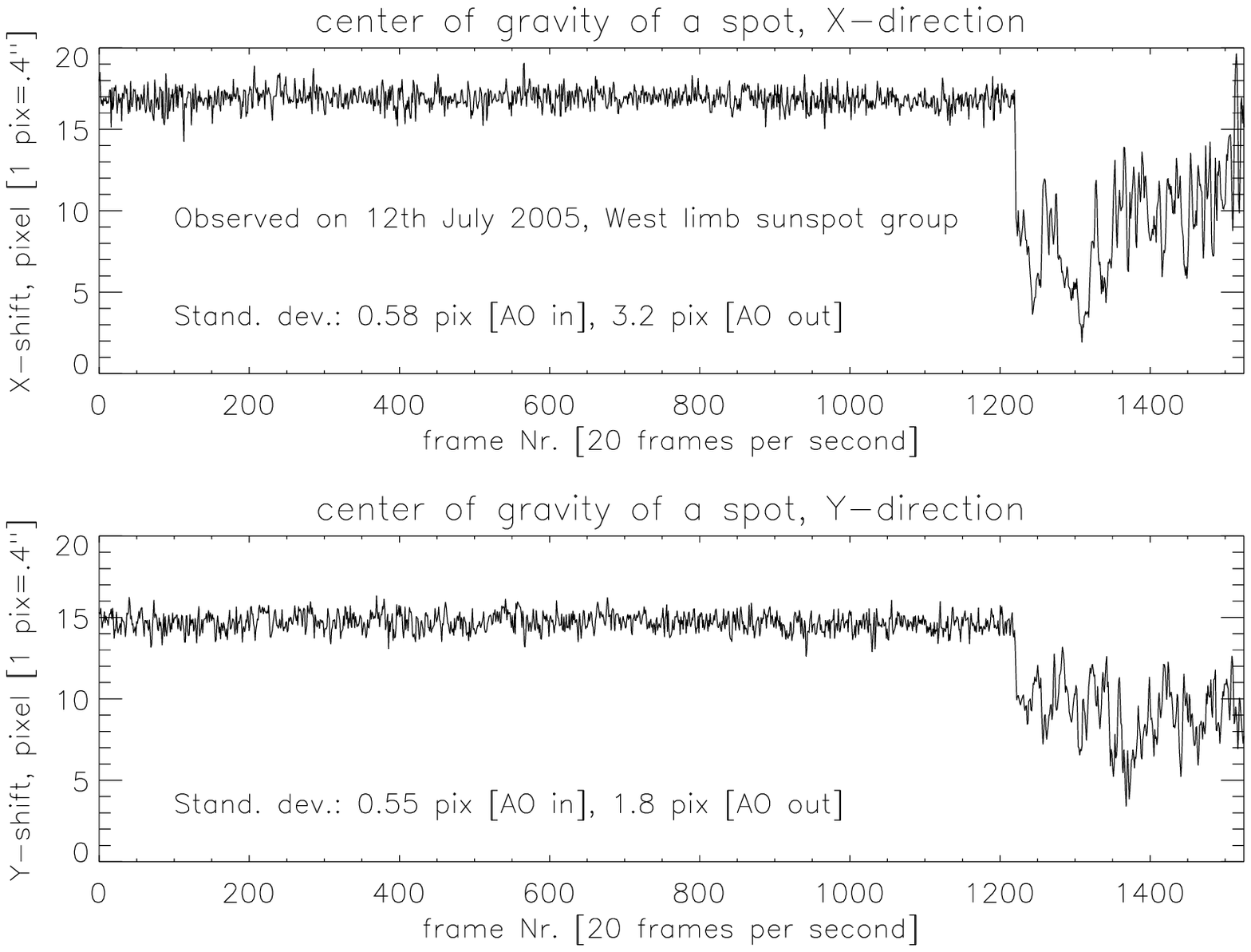,width=0.9\linewidth}}
  \caption[]{\label{ao}First tests with the tip-tilt correction of the
    AO system. The two plots show the displacement of the barycentre 
    of a sunspot in the $x$ resp. $y$ direction with and without tip-tilt
    correction. The tip-tilt mirror was switched off around frame number 1200.
}
\end{figure}

An adaptive optics (AO) system based on a tip-tilt mirror and a 37 actuator deformable 
mirror is currently being installed and tested in collaboration with the 
University of Applied Sciences of Southern Switzerland, SUPSI, and with ETH Zurich. 
The system follows the design of the infrared AO system installed at 
the McMath-Pierce Solar Telescope at Kitt Peak \cite{keller}. The first tests made with the tip-tilt
mirror have already given good results (see Figure \ref{ao}).

Two polarimeters are available. The oldest
one operating at IRSOL is the 
dual beam exchange device based on a Savart plate and rotating quarter 
and half wave retarder plates \cite{mich2}. 
A polarimetric precision of a few $10^{-4}$ can be reached, 
but it can be affected by seeing-induced cross-talk, because the technique 
requires two exposures taken at different times.
The second polarimeter is ZIMPOL \cite{hp,gand3}, which 
is installed permanently at IRSOL since 1998. Its main 
advantage is that it is free from seeing-induced effects 
thanks to its high modulation rate: 42~kHz 
(obtained with a piezoelastic modulator), or 1~kHz (obtained 
with ferro-electric liquid 
crystal modulators). Another advantage is that
the same pixels of the CCD ZIMPOL sensor
are used to measure all Stokes parameters. Therefore the Stokes $Q/I$, $U/I$ 
and $V/I$ images are not influenced by different pixel efficiencies. 
The ZIMPOL polarimetric accuracy depends mainly on
the photon statistics. With long exposure times it has already been possible
at IRSOL to reach an accuracy of about 10$^{-5}$.

\section{Scientific research projects at IRSOL}

The research projects at IRSOL take advantage of the
very good polarimetric and spectral accuracy of the instrumentation.
A large amount of observing time is available to carry out
monitoring measurements or for projects requiring long observing
times (which cannot easily be done at large telescope facilities, where
the observing time is shared by different research groups according to
a predefined program). In addition it is possible to  be  very
flexible with the programs to allow fast
reaction to particular solar events. The modular layout of the 
instrumentation inside the observing room is very convenient for
installation and testing of new instrumentation. 

\begin{figure}[htb]
  \centerline{\psfig{figure=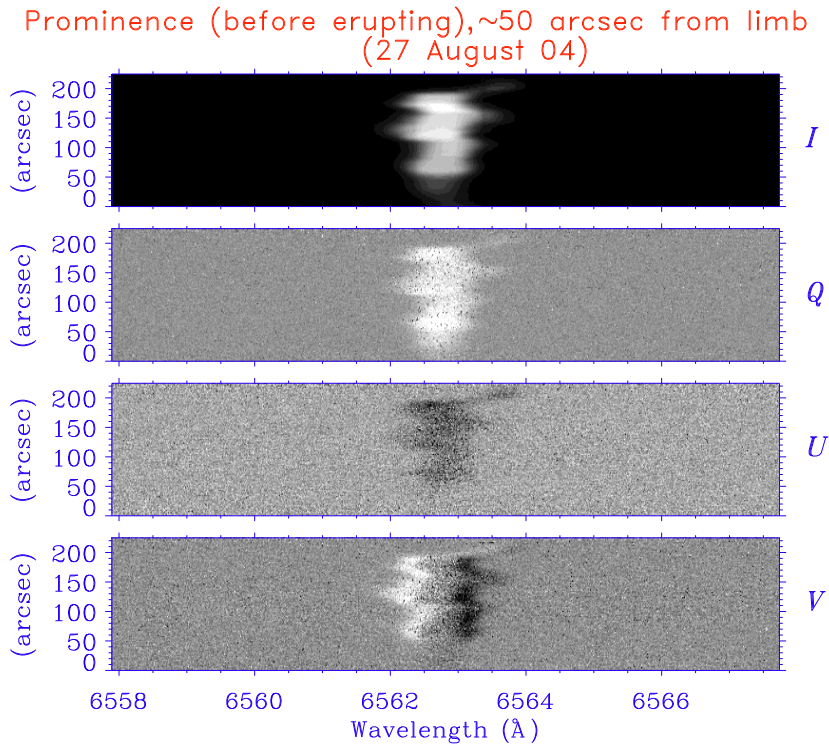,width=0.9\linewidth,angle=90}}
  \caption[]{\label{promha}Example of full Stokes spectro-polarimetric
    recording of an erupting prominence in H$_\alpha$.}
\end{figure}

Examples of the scientific results obtained at IRSOL in recent
years are:
\begin{itemize}
\item Investigations of the Hanle effect in the quiet chromosphere \cite{mich1}.
\item Publication of the first two volumes of the ``Atlas of the Second Solar Spectrum'' \cite{gand1,gand2}.
\item  Discovery of vast amounts of hidden magnetism in the solar photosphere \cite{jav,jan2}.
\item Determination of novel constraints on impact polarization in solar flares \cite{bianda05}.
\item Measurements of full Stokes profiles in prominences in
 H$_\alpha$ (Figure \ref{promha}), He D$_3$, H$_\beta$, and in spicules in He D$_3$
 \cite{renz1,ramellilindau05}.
\item First polarimetric measurements  of the Paschen-Back effect in CaH transitions \cite{sveta06}.
\end{itemize}

Different other observing programs are also foreseen in the future.
They will focus on solar magnetism and polarimetry 
with the Fabry-Perot filter system or with the spectrograph. They 
also include synoptic type programs (eg. variations of the Hanle-effect
signatures with heliographic latitude and solar cycle). Furthermore IRSOL is open to 
coordinated type programs with other observatories:
simultaneous observations of solar features with complementary sets of
instruments or supporting type observations that complement the
science of another project.

%%%%%%%%%%%%%%%%%%%%%%%%%%%%%%%%%%%%%%%%%%%%%%%% ACKNOWLEDGEMENTS
\vspace{3ex} \footnotesize \noindent 
{\bf Acknowledgements.}
We are grateful for the financial support provided by
the canton of Ticino, the city of Locarno, ETH Zurich and the
Fondazione Aldo e Cele Dacc\`o.
We also appreciate the support by the organizers of the SPSE conference.

%%%%%%%%%%%%%%%%%%%%%%%%%%%%%%%%%%%%%%%%%%%%%%%%%%% BIBLIOGRAPHY

\end{document}